\documentclass[a4paper,12pt]{article}

\usepackage{amsmath,amssymb,array,amsfonts}
\usepackage[dvipdfmx]{graphicx}
\usepackage{comment,xcolor}

\allowdisplaybreaks

\makeatletter
    
    \@addtoreset{equation}{section}
\makeatother

\usepackage[dvipdfmx]{hyperref} 
\hypersetup{
			linkcolor= blue,%
			colorlinks= true,
			citecolor = red
			}

\newcommand{\Group}[2]{{ \hbox{{\itshape{#1}}($#2$)} }}
\newcommand{\U}[1]{\Group{U\kern0.05em}{#1}}
\newcommand{\SU}[1]{\Group{SU\kern0.1em}{#1}}
\newcommand{\SL}[1]{\Group{SL\kern0.05em}{#1}}
\newcommand{\Sp}[1]{\Group{Sp\kern0.05em}{#1}}
\newcommand{\SO}[1]{\Group{SO\kern0.1em}{#1}}

\newcommand{\mybar}[1]%
    {{\kern 0.8pt\overline{\kern -0.8pt#1\kern -0.8pt}\kern 0.8pt}}
\newcommand{\sla}[1]%
    {{\raise.15ex\hbox{$/$}\kern-.57em #1}}
\newcommand{\roughly}[1]%
    {{ \mathrel{\raise.3ex\hbox{ $#1$\kern-.75em\lower1ex\hbox{$\sim$}} } }}

\newcommand{\nop}[1]{:\kern-.3em#1\kern-.3em:}

\newcommand{\al}{\ensuremath{\alpha}}
\newcommand{\be}{\ensuremath{\beta}}

\newcommand{\Ga}{\ensuremath{\Gamma}}

\newcommand{\De}{\ensuremath{\Delta}}
\newcommand{\ep}{\ensuremath{\epsilon}}

\newcommand{\ka}{\ensuremath{\kappa}}

\newcommand{\La}{\ensuremath{\Lambda}}
\newcommand{\rh}{\ensuremath{\rho}}

\newcommand{\ph}{\ensuremath{\phi}}

\newcommand{\Om}{\ensuremath{\Omega}}

\begin{document}
\setlength{\baselineskip}{18pt}
\begin{titlepage}

\begin{flushright}
KEK-TH-1735 \\
KEK-Cosmo-145 \\
arXiv:1405.4166
\end{flushright}
\vspace{10mm}
\begin{center}
{\Large\bf 
Detecting the relic gravitational wave from the electroweak phase transition at  SKA
} 
\end{center}
\vspace{15mm}

\begin{center}
{\large
Yohei Kikuta$^a$,
Kazunori Kohri$^{a, b}$
\footnote{e-mail : kohri@post.kek.jp},
Eunseong So$^{a}$
\footnote{e-mail : eunseong@post.kek.jp}
}
\end{center}
\vspace{10mm}
\centerline{{\it $^a$Department of Particle and Nuclear Physics, Graduate University for Advanced Studies }}
\centerline{{\it (Sokendai), 1-1 Oho, Tsukuba, Ibaraki 305-0801, Japan }}
\centerline{{\it $^b$KEK Theory Center, Institute of Particle and Nuclear Studies, KEK, }}
\centerline{{\it 1-1 Oho, Tsukuba, Ibaraki 305-0801, Japan }}

\vspace{0.5cm}

\begin{abstract}
  We discuss possibilities to observe stochastic gravitational wave
  backgrounds produced by the electroweak phase transition in the
  early universe. Once the first-order phase transition occurs, which
  is still predicted in a lot of theories beyond the standard model,
  collisions of nucleated vacuum bubbles and induced turbulent motions
  can become significant sources of the gravitational
  waves. Detections of such gravitational wave backgrounds are
  expected to reveal the Higgs sector physics. In particular, through
  pulsar timing experiments planned in Square Kilometre Array (SKA)
  under construction, we will be able to detect the gravitational wave
  in near future and distinguish particle physics models by comparing
  the theoretical predictions to the observations.
\end{abstract}

\end{titlepage}



\newpage

\section{Introduction}
\label{chap:intro}

Scientific research on gravitational wave is one of the most important
subjects in physics. Detecting gravitational wave directly is
essential to verify general relativity in strong gravitational fields
and explore high-energy particle physics phenomena in the early
universe. In other words, physics of gravitational wave is attractive
for both astrophysics and particle physics. Due to a weakness of its
interaction, the relic gravitational wave generated in the early
universe brings us information on the early universe for what it
was. We observe it as stochastic gravitational wave backgrounds. Quite
recently it was reported that the relic gravitational wave originated
in primordial inflation was discovered indirectly through the B-mode
polarization experiment of the Cosmic Microwave Background
(CMB)~\cite{Ade:2014xna}. Therefore direct detections of the relic
gravitational waves will take on increasing importance in the future.

In this paper, we discuss possible direct detections of the relic
gravitational wave background produced by the first-order electroweak
phase transition occurred in the early universe at around ${\cal
  O}(10^2)$~GeV.  As is well known, within the Standard Model the
effective potential of the Higgs field can not induce the first-order
phase transition unless the Higgs mass is much lighter than the
observed one~\cite{Kajantie:1993ag}. In that case no gravitational
wave is emitted because no latent heat is released during the
transition. On the other hand however, strong first-order phase
transitions are also predicted in a variety of theories beyond the
Standard Model, such as supersymmetric extended models (e.g., see
~\cite{Apreda:2001us,Huang:2014ifa}) and theories which induce a
dimensional transmutation by introducing a new scalar
field~\cite{Hambye:2013dgv} in order to explain the electroweak
symmetry breaking\footnote{Originally, models with such a strong
  first-order phase transition have been studied in terms of
  baryogenesis, e.g., see also
  ~\cite{Cohen:2012zza,Curtin:2012aa,Carena:2012np,Krizka:2012ah} with
  respect to its tension with experiments and 
  ~\cite{Pietroni:1992in,Huber:2000mg,Kang:2004pp,Menon:2004wv,Funakubo:2005pu,Huber:2006wf,Ham:2007xz,Ham:2007wc,Ashoorioon:2009nf,Chiang:2009fs,Kang:2009rd,Ahriche:2010ny,Carena:2011jy,Cheung:2012pg,Kozaczuk:2013fga,Senaha:2013kda}
  for its possible modifications. }. After the Higgs boson was
discovered~\cite{Higgs:2012gk}, we should approach various problems
related the Higgs sector in detail. Therefore, particle physicists in
the world tend to get momentum to tackle the physics at the
electroweak phase transition head-on.

Investigations of the Higgs sector by using gravitational wave
experiments are indeed exciting since we can explore particle physics
through observations at cosmological scales. This kind of the
verification for the Higgs sector is complementary to experiments that
directly explore the theories beyond the Standard Model like the Large
Hadron Collider (LHC) experiments and can be even much more powerful
in some ways.

Since various experiments are planned to try to observe the
gravitational waves, they cover a wide range of frequencies
$10^{-9}$~Hz $ \lesssim f \lesssim 10^3$~Hz. In principle future
experiments such as eLISA~\cite{eLISA} and
DECIGO/BBO~\cite{Seto:2001qf,Kudoh:2005as,Alabidi:2013lya} have been
known to detect the relic gravitational waves produced by the
electroweak phase transition in future for the frequencies
$10^{-7}$~Hz $ \lesssim f \lesssim 10$~Hz. In this paper, we further
discuss possibilities to observe the relic gravitational waves through
the pulsar timing experiments at Square Kilometre Array (SKA) under
construction for the frequencies $10^{-9}$~Hz $ \lesssim f \lesssim
10^{-4}$~Hz~\cite{Kramer:2010}. The phase 1 and the phase 2 of SKA
will starts from 2018 and 2023, respectively~\cite{SKAweb}.

In addition, so far effects by a large vacuum energy at a false vacuum
on the phase transition has not been well examined. In this paper, we
study the effect of the finite vacuum energy at the false vacuum in
terms of cosmology.

This paper is organized as follows. In Section 2 we show model
independent analyses of gravitational wave produced by the first-order
electroweak phase transition. Section 3 is devoted to study the effect
of the vacuum energy at the false vacuum. In Section 4, we show the
experimental detectabilities of the relic gravitational wave
background. Finally, in Section 5 we summarize our works.

\section{Model-independent analysis}
\label{sec:model_independent}

When the first-order phase transition occurs, the universe make a
transition from a false vacuum state to a true vacuum state. There
exists an effective potential barrier between the true and the false
vacua. Then, the transition occurs due to thermal fluctuations and a
quantum tunneling effect. In other words, the true vacuum bubbles are
produced inside the false vacuum state. However, the bubble nucleation
itself does not generate any gravitational waves because of its
spherical symmetric nature. The spherical symmetry is broken when they
collide through their expansion, generating stochastic gravitational
waves~\cite{Kehayias:2009tn}.  Fine details of the colliding regions
are not so important to calculate the gravitational wave
production. However, the gravitational wave is rather dominated by the
gross features of the evolving bubble, which depends on kinetic
energies of uncollided bubble
walls~\cite{Kosowsky:1991ua,Caprini:2007xq}.  These facts mean that
so-called ``the envelope approximation'' should be a good
approximation for evaluating the amount of the produced gravitational
wave signals~\cite{Kosowsky:1992vn}\footnote{On the other hand, see
  also a recent criticism reported by~\cite{Hindmarsh:2013xza}.}.

In addition, the bubble expansion causes a macroscopic motion of
cosmic plasma. When the bubbles collide, turbulence occurs in the
fluid, which can become a significant source of the gravitational wave
background~\cite{Grojean:2006bp,Caprini:2009yp}.

In this section, we introduce analytical methods to study the
gravitational waves produced by the first-order phase transition. We
take two most important parameters, $\alpha$ and $\tilde{\beta}$,
characterizing the gravitational waves from the first-order phase
transition. Then we show that general model parameters sufficiently
reduce to only those two parameters when we discuss signals of the
relic gravitational wave background.

\subsection{Basics}

We adopt definitions of parameters used in this section mainly by
following the ones in Ref.~\cite{Grojean:2006bp}. We discuss phenomena
on the basis of the Friedman-Robertson-Walker universe, in which
$a(t)$ represents scale factor of the universe. We assume that the
phase transition occurs at a cosmic temperature $T_*$ which is the
order of ${\cal O}(10^2)$~GeV. The gravitational wave of the frequency
$f_*$ has arrived to us to be the present frequency~$f$.  Hereafter
the subscript ``$*$'' denotes a physical quantity at the phase
transition.  Then, the frequency we currently observe is represented
by
\begin{align}
  f = f_* \frac{a_*}{a_0} = f_* \left( \frac{g_{s0}}{g_{s*}} \right)^{1/3} \frac{T_0}{T_*},
\end{align} 
where the subscript ``$0$'' means a value at the present.  Here we used
the adiabatic expansion of the universe (i.e., the entropy $S \propto
a^3 g_s (T) T^3 = $ const ). $g_s$ means the effective degrees of
freedom,
\begin{align}
g_s(T)=\sum_{\rm boson}g_i(\frac{T_i}{T})^3+\frac{7}{8}\sum_{\rm fermion}g_i(\frac{T_i}{T})^3
\end{align}
where $g_i$ counts the internal degrees of freedom of $i$-th particle. In
the current universe, we have $g_s(T_0=2.725K)\simeq 3.91$. In terms of
Hubble parameter, the frequency is given by
\begin{align}
  f \simeq 6 \times 10^{-3} \left( \frac{g_*}{100} \right)^{1/6} \frac{T_*}{100 \text{GeV}} \frac{f_*}{H_*}~\text{mHz},
\end{align} 
where $g_* = g_s$ for $T \gg$ 1 MeV.  Therefore we expect the typical
frequency for the gravitational wave produced at the electroweak phase
transition to be at around $\sim 10^{-3}$ mHz -- $10^{-2}$ mHz.

The energy density of the stochastic gravitational wave 
background~\footnote{It is related with the strain $\sqrt{S_{\rm GW}}$ to be $\Om_{\rm{GW}} h^2 = 3.132 \times 10^{35} (f/{\rm Hz})^3 (\sqrt{S_{\rm GW}} / {\rm Hz}^{-1/2})^2$.
}
is calculated to be
\begin{align}
  \Om_{\rm{GW}} h^2 \equiv \frac{\rh_{\rm{GW}}}{\rh_{c}} h^2 = \Om_{\rm{GW}*} h^2 \left( \frac{a_*}{a_0} \right)^4  \left( \frac{H_*}{H_0} \right)^2
  	\simeq 1.67 \times 10^{-5}h^{-2} \left( \frac{100}{g_*} \right)^{1/3} \Om_{\rm GW*}.
\end{align} 
where we used $ \rh_{GW} a_0^4 = \rh_{GW*} a_*^4 ,  \ \  \rh_c H_0^{- 2} = \rh_{c*} H_*^{- 2}, \ \  H_0 = 2.1332 \times h \times 10^{-42} \text{GeV},$
with $H$ Hubble parameter and $h$ its reduced value. The subscript ``$0$''
denotes the value at the current epoch. In the next section we will
show how we can calculate $\Om_{\rm GW} h^2= \Om_{\rm coll}h^2 +
\Om_{\rm turb}h^2$ in terms of two fundamental parameters ($\alpha$ and
$\tilde{\beta}$), which is the summation of the two contributions from
the bubble collision ($\Om_{\rm coll}h^2$) and the turbulence
($\Om_{\rm turb}h^2$).

\subsection{Fundamental parameters, $\alpha$ and  $\tilde{\beta}$}

We introduce two important parameters $\al$ and $\tilde{\beta}$ to
discuss model-independent analyses.  At a finite temperature, the
bubble nucleation rate of the phase transition is represented
by~\cite{Grojean:2006bp}
\begin{align}
  \Ga (T) = \Ga_0 (T) e^{- S(T)} \simeq \Ga_0 (T) e^{- \frac{S_3}{T}},
\label{eq:bubble_nucleation}
\end{align} 
where $\Ga_0(T)$ has units of energy to the fourth power and is typically
represented by $ \Ga_0 (T) \sim T^4$. $S_3$ stands for the euclidean
action of the system~\cite{Linde:1981zj,Sagunski:2012ufa},
\begin{align}
  S_3(T) = \int 4 \pi r^2 \left[ \frac{1}{2} \left( \frac{d \ph_b}{d r} \right)^2 + V_{\rm{eff}} (\ph_b, T) \right].
\end{align} 
Notice that $S_3$ becomes time-independent at a high
temperature~\cite{Sagunski:2012ufa}. $V_{\rm{eff}}(\phi,T)$ means the
effective potential of the field $\phi$ at a finite temperature $T$.
$\ph_b$ represents a bubble profile of the field $\phi$.  $r$
denotes a radius in the polar coordinates. Then the bubble profile
is obtained by solving the bounce equation,
\begin{eqnarray}
  \label{eq:bounce0}
  \frac{d^2 \ph_b}{d r^2} + \frac{2}{r} \frac{d \ph_b}{d r} - \frac{ \partial V}{ \partial \ph_b} = 0,
\end{eqnarray}
with
\begin{eqnarray}
  \label{eq:dphidr}
\left. \frac{d \ph_b}{d r} \right|_{r=0} = 0,
\end{eqnarray}
and
\begin{eqnarray}
  \label{eq:phib0}
        \left. \ph_b \right|_{r = \infty} = 0.
\end{eqnarray}
Since the bubble nucleation rate has an exponential dependence, a key
is a behavior of $S_3 / T$. By taking the time derivative of the
action, we define
\begin{align}
  \be \equiv \left. - \frac{ d S }{ d t } \right|_{t_*}.
\end{align} 
In a neighborhood of $t_*$, we naturally expect a series expansion to
be $S (t) = S(t_*) - \be (t - t_*) + \dots$ Here we introduce a
dimension-less parameter to express the time derivative of the action,
\begin{align}
 \tilde{\beta}\equiv \frac{\be}{H_*} = T_* \left. \frac{d S}{d T} \right|_{T_*} = T_* \frac{d}{d T} \left. \left( \frac{S_3}{T} \right) \right|_{T_*},
\end{align} 
where we used a property of the adiabatic expansion of the universe,
$dT / dt = - T H$. This $\tilde{\beta}$ is one of the most important
parameters to characterize the shape of the gravitational wave
spectrum.  It is sufficient to look at the relationship used to
determine the typical value of this, being able to percolate properly
even for the exponentially-expanding universe\footnote{There are also
  another evaluations such as $ \int dt \frac{\Ga}{H^3} \sim {\cal O}
  (1)$ appeared in other works.  However, we have checked that this
  difference does not change our conclusion.}.
\begin{align}
  \frac{\Ga}{H^4} \sim {\cal O} (1).
\end{align} 
Using this condition, it is possible to estimate the value of $\tilde{\beta}$ for each model.

Another important parameter is a quantity that represents how much
latent heat is released at the phase transition.  In the symmetry
phase, we denote the false vacuum energy density and the thermal
energy density to be $\ep(T)$, and $\rh_{\rm{rad}}(T)$,
respectively. Then, the parameter $\al = \al (T)$ is defined by
\begin{align}
  \al \equiv \frac{\ep(T)}{\rh_{\rm{rad}}(T)}.
\end{align} 
Here, the energy density of the false vacuum  is represented by
\begin{align}
  \ep \equiv \De V_{\rm{eff}} - T \De s =\De V_{\rm{eff}} - T \frac{\partial \De V_{\rm{eff}}}{\partial T},
\end{align} 
where
\begin{eqnarray}
  \label{eq:DeltaVeff}
  \De V_{\rm{eff}}= \De V_{\rm{eff}}(T) \equiv V_{\rm{eff}}(\phi_{\rm false},T)  -  V_{\rm{eff}}(\phi_{\rm true},T),
\end{eqnarray}
with $\phi_{\rm true}$ and $\phi_{\rm false}$ being the field values at
the true and false vacua, respectively.  Also, the energy density of radiation
$\rh_{\rm{rad}}$ is given by
\begin{align}
\rh_{\rm{rad}}(T)=\frac{\pi^2}{30}g_*T^4.
\end{align}

Using those two parameters ($\alpha$ and $\tilde{\beta}$), a peak
spectrum of the gravitational wave $\tilde{\Omega}h^2$ at a peak
frequency $\tilde{f}$ is represented by~\cite{Kamionkowski:1993fg}
\begin{align}
  \tilde{f}_{\rm{coll}} &\simeq 5.2 \times 10^{-3} \frac{\be}{H_*} \frac{T_*}{ 100 \text{GeV} } \left( \frac{g_*}{100} \right)^{1/6} \text{mHz}, \\
  \tilde{\Om}_{\rm{coll}} h^2  &\simeq 1.1 \times 10^{-6}\ka^2 \left( \frac{H_*}{\be} \right)^2 
  	\left( \frac{\al}{1+\al} \right)^2 \frac{v_b^3}{ 0.24 + v_b^3 } \left( \frac{100}{g_*} \right)^{1/3}, \\
  \tilde{f}_{\rm{turb}} &\simeq 3.4 \times 10^{-3}  \frac{u_s}{v_b} \frac{\be}{H_*} \frac{T_*}{ 100 \text{GeV} } 
  	\left( \frac{g_*}{100} \right)^{1/6}  \text{mHz}, \\
  \tilde{\Om}_{\rm{turb}} h^2  &\simeq 1.4 \times 10^{-4} u_s^5 v_b^2 \left( \frac{H_*}{\be} \right)^2 \left( \frac{100}{g_*} \right)^{1/3}.
\end{align} 
Here the subscript ``coll'' and ``turb'' denote the values in cases of
the bubble collision and the turbulence, respectively. In these
expressions, the bubble velocity $v_b$, the fluid velocity $u_s$ and
the efficiency factor $\kappa$ are expressed 
as a function of $\al$ to be~\cite{Kamionkowski:1993fg}
\begin{align}
  v_b (\al) &= \frac{ \frac{1}{\sqrt{3}} + \sqrt{ \al^2 + \frac{2 \al}{3} } }{1+\al}, \\
  u_s (\al) &= \sqrt{ \frac{ \ka \al }{ \frac{4}{3} + \ka \al } }, \\
  \ka (\al) &= \frac{1}{ 1 + 0.715 \al } \left( 0.715 \al + \frac{4}{27} \sqrt{ \frac{3 \al}{2} }  \right).
\end{align} 

In case of the bubble collision, the entire spectrum has been also
calculated analytically. Using an envelope approximation, the full
spectrum of the gravitational wave from the bubble collision is given
by~\cite{Huber:2008hg}
\begin{eqnarray}
  \label{eq:tailColl}
  \Om_{\rm{coll}}(f)h^2=\tilde{\Om}_{\rm{coll}} h^2
  \frac{(a+b)\tilde{f}^bf^a}{b\tilde{f}^{a+b}+a{f}^{a+b}},
\end{eqnarray}
where the value of $a$ and $b$ lie in the range $a\in [2.66,2.82]$,
and $b\in [0.90,1.19]$. In case of the strong first-order phase
transition, by a numerical simulation using a large number of
colliding bubbles, the authors of \cite{Huber:2008hg} obtained
$a\simeq 2.8$, and $b\simeq 1$. In Eq.~(\ref{eq:tailColl}) it is
easily found that $\Omega_{\rm coll} h^2 (f)= \tilde{\Omega}_{\rm
  coll}h^2 $ at $f = \tilde{f}$.
There is a remark that the formulae given here are available only when
$\tilde{\beta}$ is sufficiently large~\cite{Huber:2008hg}. 

As will be shown later, the effects due to the tails parts of the
spectrum given in Eq.(\ref{eq:tailColl}) on experimental
detectabilities are quite small. Hence, even if we do not adopt full
expressions for the spectrum in the turbulent case, which has not been
known analytically, our results should not change significantly only
in the current purposes. Therefore we may take $\Omega_{\rm turb}h^2
(f) = \tilde{\Omega}_{\rm turb}h^2 (f) $ for any $f$'s approximately
as a full spectrum for the turbulent case.

\section{Effects of the vacuum energy at the false vacuum}
\label{sec:effect_of_vacuum} 

In the previous section, we adopted the parametrizations, in which we
took a limit that the vacuum energy is completely negligible. However,
here we carefully check possible effects on the productions of the
relic gravitational wave background.

\begin{figure}[htbp]
\begin{center}
  \includegraphics[scale=0.7]{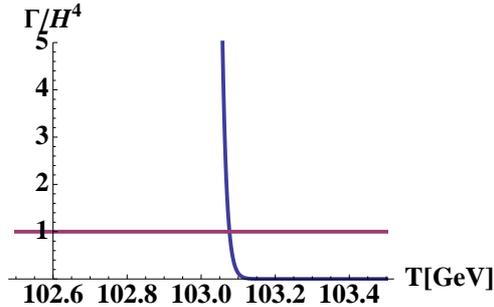}
  \caption{Plot of the condition for completion of the percolation
    $\Ga/ H^4 = 1$. The horizontal axis is the cosmic temperature in
    GeV. Here we took vacuum energy $\La_{\rm{vac}} \equiv \Delta
    V_{\rm eff}(T=0)=0$. }
\label{fig:percolation_novacuum}
\end{center}
\end{figure}

First, we investigate how the percolation is influenced by the vacuum
energy.  Here we consider typical cases in the Minimal Supersymmetric
Standard Model (MSSM) as a specific example (see
Appendix~\ref{app:susy} for the details). If we ignore the vacuum
energy, we have obtained $T_* \sim 103.1 \text{GeV}$ in order to
complete the percolation as is shown in
Fig.~\ref{fig:percolation_novacuum}.

\begin{figure}[htbp]
	\begin{center}
		\includegraphics[scale=0.7]{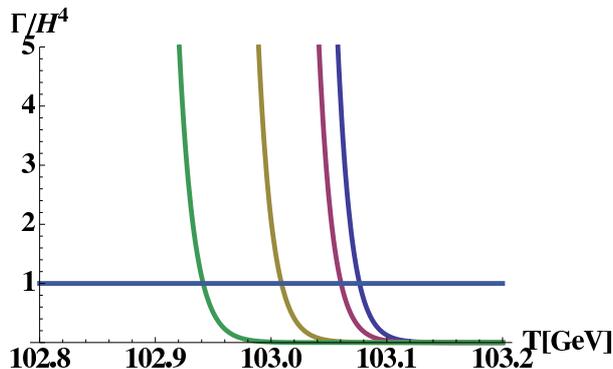}
                \caption{Same as Fig.~\ref{fig:percolation_novacuum},
                  but changing the vacuum energy.  From right to left,
                  we took $\La_{\rm{vac}} \equiv \Delta
                  V_{\rm eff}(T=0)=$ $(0)^4, (250)^4, (500)^4$, and
                  $(1000)^4 \text{GeV}^4$.  }
		\label{fig:percolation_vacuum}
  	\end{center}
\end{figure}

Next we incorporate the effect of the vacuum energy in this setup. For
a concrete calculation, we calculate $\Ga/H^4$ by parameterizing the
vacuum energy $\Lambda_{\rm vac} \equiv \Delta V_{\rm eff}(T=0)$. As
seen in the previous section, the bubble nucleation is determined by
the value of the euclidean action. However, there is no effect from
the vacuum energy on the nucleation because the constant term is
renormalized in the definition of the nucleation rate. Only the Hubble
parameter, $H^2 = \rho / (3 m_{\rm pl}^2)$ with $m_{\rm pl}$ Planck
mass, should be changed by adding the vacuum energy to the total
energy density $\rho = V_{\rm eff}(\phi,T) + \rho_{\rm rad}$. The
result is plotted in Fig.~\ref{fig:percolation_vacuum}. As seen in
this figure, the effect is only a mild change on $T_* $ with a small
difference by the order of ${\cal O}(0.1) \text{GeV} $. That is
because the bubble nucleation has the exponential dependence on $T$,
which is the dominant contribution to possibly change $\Gamma/H^4$.
\footnote{As was mentioned in the previous footnote, there is another
  evaluation $\int dt \frac{\Ga}{H^3} \sim 1$.  We have also
  calculated transition temperature in those two cases, i.e., no
  vacuum energy $\Lambda_{\rm vac}=0$, and $\Lambda_{\rm vac}=(500)^4
  \text{GeV}^4$.  In those cases the corresponding transition
  temperatures are $101.91 \text{GeV}$ and $101.73 \text{GeV}$,
  respectively. Therefore, although there seems to exist a small
  difference between $\Gamma/H^4$ and $\int dt \frac{\Ga}{H^3}$, it
  does not change our conclusions so much.}

\section{Detectability of relic gravitational wave}
\label{sec:experimental_test} 

In this section, we discuss detectabilities of the relic gravitational
wave background produced at the electroweak phase transition by using
two fundamental parameters $\alpha$ and $\tilde{\beta}$. In case of the phase transition at the
electroweak scale, the useful experiments should be
eLISA~\cite{eLISA}, Ultimate
DECIGO~\cite{Seto:2001qf,Kudoh:2005as,Alabidi:2013lya} and
SKA~\cite{Kramer:2010}. The sensitivities of the experiments are
summarized in Refs.~\cite{Alabidi:2013lya,Regimbau:2012ir}
 
There is also a limit from non-detections of extra radiation through
the CMB observation such as the Planck satellite experiments as an
additional constraint.  The extra radiation like the stochastic
gravitational background can be measured as a deviation of the
effective number of the neutrino species $N_{\nu, {\rm eff}}$ from
three to be $\Delta N_{\nu} = N_{\nu, {\rm eff}} - 3$. Then the energy
fraction of the extra radiation at present can be expressed by
$\Omega_{\rm extra}h^2 = 5.108 \times 10^{-6 } \Delta N_{\nu}$.  So
far the Planck collaborations have reported that observationally they
had an upper bound on $\Delta N_{\nu,}$ to be $\Delta N_{\nu,}
\lesssim 1$.~\cite{Ade:2013zuv}. Then we obtain an upper bound on the
energy fraction of the relic gravitational wave background,
\begin{eqnarray}
  \label{eq:OmegaDeltaNnu}
  \Omega_{\rm GW}h^2 < 5.108 \times 10^{-6 } 
\left( \frac{\Delta N_{\nu}}{1} \right).
\end{eqnarray}
This is effective over a broad range of frequencies for $10^{-17} {\rm
  Hz}\lesssim f$,~\footnote{Or the range is represented in terms of
  the comoving wave number to be $10^{-2}$~Mpc$^{-1} \lesssim k$
  through $f = 1.535 \times 10^{-5} {\rm Hz} (k /10^{10} {\rm
    Mpc^{-1}})$.}  which is wider than the one obtained by
big-bang nucleosynthesis (BBN).

\begin{figure}[htbp]
\begin{center}
\includegraphics[scale=0.55]{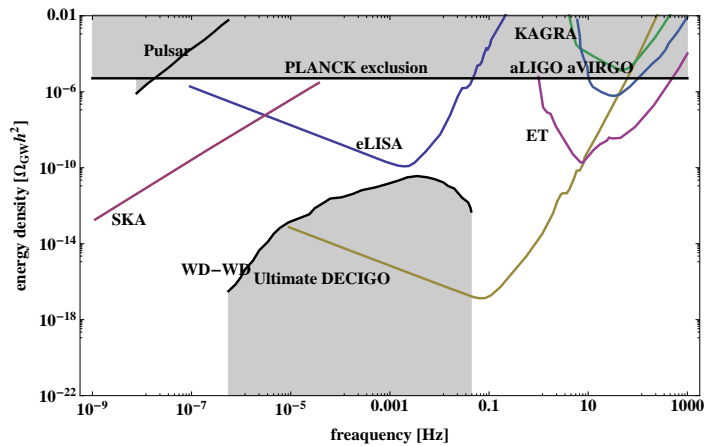}
\caption{ Experimental sensitivities of eLISA, DECIGO, SKA, Advanced
  LIGO/VIRGO, KAGRA, and ET. The horizontal line means the upper bound
   from the CMB observations by PLANCK given in
  Eq.~(\ref{eq:OmegaDeltaNnu}). ``Pulsar'' denotes the upper bound
  obtained from the existing pulsar timing experiments. The WD-WD line
  stands for the foreground noise from white dwarf
  binaries~\cite{Schneider:2010ks}. The detail of each experimental
  line is given in the text and
  Refs.~\cite{Alabidi:2013lya,Regimbau:2012ir}}
\label{fig:experimental_sensitivity}
\end{center}
\end{figure}

We calculate the spectra by changing the parameters to be $\al =
[10^{-1}, 10]$, and $\tilde{\be} = [10^{-1}, 10^4] $ with the
transition temperature $T_*=$ 70 GeV and 100 GeV with $g_*=106.75$.

\begin{figure}[htbp]
\begin{center}
\includegraphics[scale=0.55]{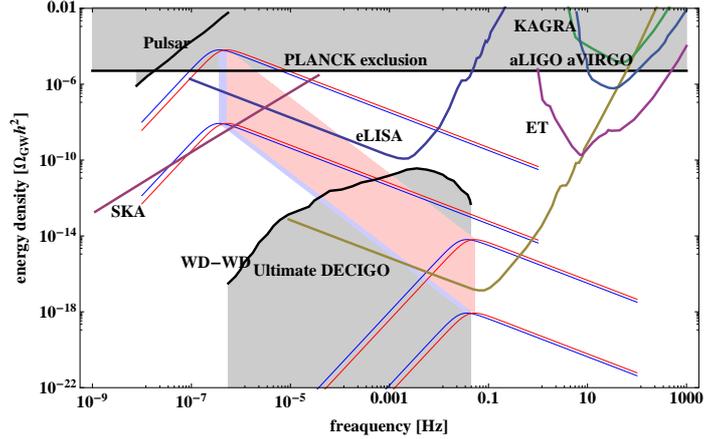}
\caption{Signals of the relic gravitational wave background in case of
  the bubble collision. The band regions mean the peak signals
  $\tilde{\Omega}h^2$ for $T = 70 \text{GeV}$, and $T = 100
  \text{GeV}$ from the left to the right, respectively. The broken
  power means the corresponding full spectrum whose peak is located at
  $\Omega h^2=\tilde{\Omega}h^2$. The model parameters are changed to
  be
  $\{\al,\tilde{\be}\}=\{0.1,0.1\},\{0.1,10^4\},\{10,0.1\},\{10,10^4\}$. We
  assumed $g_* = 106.75$.}
\label{fig:bubble_collision}
\end{center}
\end{figure}
In case of the bubble collision, we plot the obtained signals in
Fig.~\ref{fig:bubble_collision}. The peak frequency is controlled only
by $\tilde{\beta}$.  On the other hand, the peak signal is determined
by both $\alpha$ and $\tilde{\beta}$. There exist regions which have
been already excluded by the Planck constraint
[Eq.~(\ref{eq:OmegaDeltaNnu})]. It is remarkable that there are
parameter regions, which only SKA can observe at a small $\tilde{\beta}$.

\begin{figure}[htbp]
\begin{center}
\includegraphics[scale=0.55]{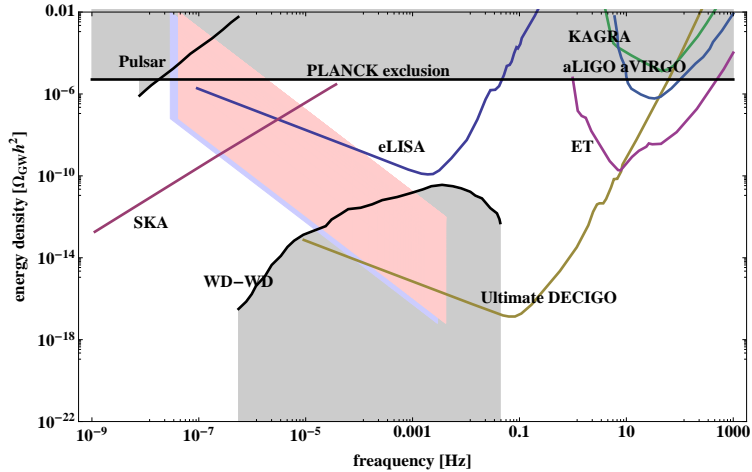}
\caption{Same as Fig.~\ref{fig:bubble_collision}, but for the case of the turbulence.}
\label{fig:turbulence}
\end{center}
\end{figure}

In Fig.~\ref{fig:turbulence}, we plot the signals of the relic
gravitational wave background sourced by the turbulence.  Contrary to
the case of the bubble collision, it is notable that the peak
frequency depends on both $\alpha$ and $\tilde{\beta}$. Of course, the
peak signal is also determined by both $\alpha$ and
$\tilde{\beta}$. The turbulence makes an important contribution
to the signal and has larger detectable parameter regions\footnote{By recent hydrodynamic simulations, e.g., \cite{Hindmarsh:2013xza,Giblin:2014qia,Espinosa:2010hh}, it was pointed out that its contribution might be smaller.}.

\begin{figure}[htbp]
\begin{center}
\includegraphics[scale=0.4]{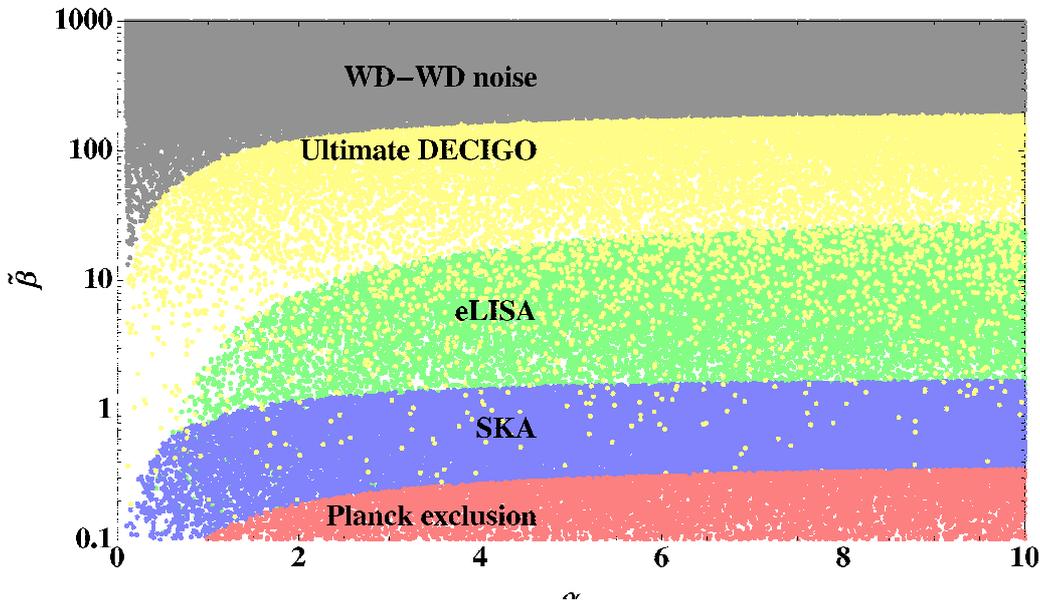}
\caption{ Detectabilities in the (${ \al, \tilde{\be}}$) plane for the
  signals sourced by the bubble collision. From the top to the bottom,
  the cases for Ultimate DECIGO, eLISA, SKA are plotted, respectively.
  The excluded regions by the Plank constraint
  (Eq.~\ref{eq:OmegaDeltaNnu}) are also plotted. The WD-WD noise means
  the region where signals are covered by the foreground noise by the
  WD-WD binaries.}
\label{fig:plane_collision}
\end{center}
\end{figure}

\begin{figure}[htbp]
\begin{center}
\includegraphics[scale=0.4]{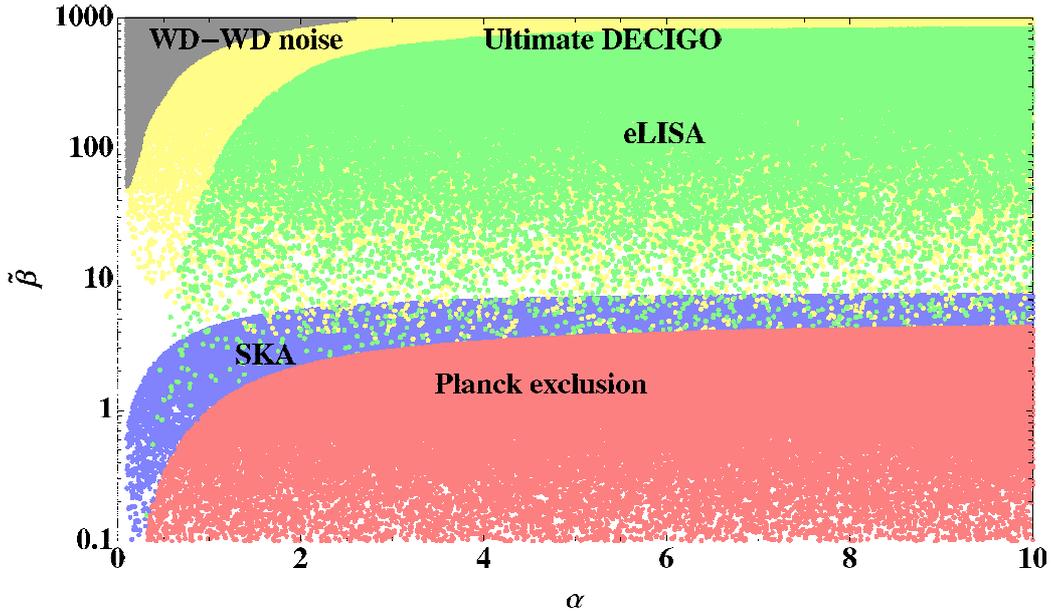}
\caption{
Same as Fig.~\ref{fig:plane_collision}, but for the signals sourced by the turbulence.
}
\label{fig:plane_turbulence}
\end{center}
\end{figure}

We scanned parameter regions in terms of detectabilities in the (${
  \al, \tilde{\be}}$) plane. In Fig.~\ref{fig:plane_collision} the
case of the bubble collision is plotted.  Here, we consider only the
case of $T_*=100 \rm{GeV}$. It displays the three regions that can be
detected in three experiments (Ultimate DECIGO, eLISA and SKA). The
excluded regions by the Plank [Eq.~(\ref{eq:OmegaDeltaNnu})] is also
plotted at the bottom. Top regions are covered by the WD-WD noise.  We
have checked that the allowed region does not change much even if we
consider the corresponding tail of the full spectrum shown in
Eq.~(\ref{eq:tailColl}). In Fig.~\ref{fig:plane_turbulence}, we also
plot the case of the turbulence.

%

\section{Conclusions}
\label{sec:conclusions} 

After the discovery of the Higgs boson, particle physicists in the
world tend to get momentum to tackle the physics at the electroweak
phase transition head-on and approach a various serious problems
related the Higgs sector in detail.  Therefore, it is attractive to
revisit a variety of possible scenarios for the electroweak phase
transition. We have examined detectabilities of the stochastic
gravitational wave background produced by the first-order electroweak
phase transition in a general setup with carefully considering effects
of the vacuum energy on the expansion of the universe.

We have shown that the relic gravitational wave background
produced at the electroweak epoch will be observed by the future
experiments, such as SKA, eLISA and DECIGO. In particular, the small
$\tilde{\beta}$ regions, which is naturally predicted in some particle
physics models such as MSSM, will be able to be searched by SKA very
near future.

\subsection*{Acknowledgments} 

We would like to thank Kenta Hotokezaka, Kunihito Ioka,
and Eibun Senaha for useful discussions.  This work is partially
supported by the Grant-in-Aid for Scientific research from the
Ministry of Education, Science, Sports, and Culture, Japan,
Nos. 25.2309 (Y.K), and 21111006, 22244030, 23540327, 26105520
(K.K.). The work of K.K.  is also supported by the Center for the
Promotion of Integrated Science (CPIS) of Sokendai
(1HB5804100).

\appendix
\label{sec:appendix}

\section{Scalar sector in  MSSM}
\label{app:susy}
In this section we show details of the adopted models which are
motivated by MSSM~\cite{Apreda:2001us}. Those models are used to
calculate specific physical variables in
Sec~\ref{sec:effect_of_vacuum}. We refer only a scalar potential
required here to be $V(\phi, T) = V_{\rm{eff}}(\phi, T)$ with a
constant vacuum energy $\Lambda_{\rm vac} = \Delta V_{\rm
  eff}(T=0)$. The effective potential $V_{\rm{eff}}(\phi, T)$ is then
represented by~\cite{Apreda:2001us}
\begin{align}
  V_{\rm{eff}}(\phi, T) = V_0 (\phi) + V_1 (\phi, T) + V_2 (\phi, T). 
\end{align}
The tree level potential $V_0 (\phi)$ depends only on the scalar
field.
\begin{align}
  V_0 (\phi) = - \frac{m_H^2}{4} \phi^2 + \frac{m_H^2}{v^2} \phi^4,
\end{align}
with $m_H$ being the Higgs mass and $v$ being the vev of the Higgs
field.

$V_1 (\phi, T)$ is calculated at the thermal one-loop level. The
following is a result of the high temperature expansion,
\begin{eqnarray}
  \label{eq:V1}
  V_1 (\phi, T) &=& \frac{T^2}{2 v^2} \left( \frac{m_H^2}{4} + \frac{5 m_W^2}{6} + \frac{5 m_Z^2}{12} + m_t^2 \right) \phi^2 \nonumber \\
   &&- T \left( E_{\rm SM} + 2 N_c 
     \frac{ ( m_{\rm stop}^2 + \Pi_{\rm stop} )^{3/2} }{12 \pi} \right)
   \phi^3,
\end{eqnarray}
where the number of colors $N_c=3$, with $m_t$, $m_{\rm stop}$, $m_W$
and $m_Z$ being masses of the top quark, the scalar top quark, the $W$
boson and the $Z$ boson, respectively.  Here we introduced
\begin{align}
   E_{\rm SM} &= \frac{1}{3} \left( \frac{ 2 m_W^3 + m_Z^3 }{2 \pi v^3} \right), \\
   m_{\rm stop}^2 &= - m_U^2 + \left( 0.15 \frac{m_Z^2}{v^2} \cos (2 \be_{\rm MSSM}) + \frac{m_t^2}{v^2} \right) \phi^2 , \\
   \Pi_{\rm stop} &= \frac{4 g_s^2}{9} T^2 + \frac{h_t^2}{6} (1 + \sin^2 {\be_{\rm MSSM}}) T^2 + \left( \frac{1}{3} - \frac{1}{18} |\cos (2 \be_{\rm MSSM})| \right) g'^2 T^2,
\end{align}
where $g'$ and $g$ are the $U(1)_Y \times SU(2)_L$ gauge
coupling constants, $g_s$ is the strong gauge coupling constant, $h_t$
is the top Yukawa coupling, $m_U^2$ is the model parameter of the
soft-mass squared, and $\be_{\rm MSSM}$ is an angle defined by
$\tan\be_{\rm MSSM}$ to be the ratio of two Higgs' vevs.  Finally,
$V_2 (\phi, T)$ is calculated by the two loop effect by incorporating
the effect of weak boson and scalar top quark (stop),
\begin{align}
  V_2 (\phi, T) = \frac{ \phi^2 T^2 }{32 \pi^2} \left( \frac{51 g^2}{16} - 3 h_t^4 \sin^4 (\be_{\rm MSSM}) + 8 g_s^2 h_t^2 \sin^2 (\be_{\rm MSSM}) \right) \log \left( \frac{\La}{\phi} \right).
\end{align}



\end{document}